\documentclass[11pt]{amsart}
\usepackage{amssymb,amsmath,mathrsfs,epsfig} 
\evensidemargin \oddsidemargin
\newtheorem{theorem}{Theorem}

\newcommand{\remove}[1]{}
\newcommand\nc\newcommand
\newcommand\Proof{\noindent{\sc Proof. }}

\nc\bfa{{\bf a}}\nc\bfA{{\bf A}}\nc\cA{{\mathcal A}}
\nc\bfb{{\bf b}}\nc\bfB{{\bf B}}\nc\cB{{\mathcal B}}
\nc\bfc{{\bf c}}\nc\bfC{{\bf C}}\nc\cC{{\mathcal C}}
\nc\bfd{{\bf d}}\nc\bfD{{\bf D}}\nc\cD{{\mathcal D}}
\nc\bfe{{\bf e}}\nc\bfE{{\bf E}}\nc\cE{{\mathcal E}}
\nc\bff{{\bf f}}\nc\bfF{{\bf F}}\nc\cF{{\mathcal F}}
\nc\bfg{{\bf g}}\nc\bfG{{\bf G}}\nc\cG{{\mathcal G}}
\nc\bfh{{\bf h}}\nc\bfH{{\bf H}}\nc\cH{{\mathcal H}}
\nc\bfi{{\bf i}}\nc\bfI{{\bf I}}\nc\cI{{\mathcal I}}
\nc\bfj{{\bf j}}\nc\bfJ{{\bf J}}\nc\cJ{{\mathcal J}}
\nc\bfk{{\bf k}}\nc\bfK{{\bf K}}\nc\cK{{\mathcal K}}
\nc\bfl{{\bf l}}\nc\bfL{{\bf L}}\nc\cL{{\mathcal L}}
\nc\bfm{{\bf m}}\nc\bfM{{\bf M}}\nc\cM{{\mathcal M}}
\nc\bfn{{\bf n}}\nc\bfN{{\bf N}}\nc\cN{{\mathcal N}}
\nc\bfo{{\bf o}}\nc\bfO{{\bf O}}\nc\cO{{\mathcal O}}
\nc\bfp{{\bf p}}\nc\bfP{{\bf P}}\nc\cP{{\mathcal P}}
\nc\bfq{{\bf q}}\nc\bfQ{{\bf Q}}\nc\cQ{{\mathcal Q}}
\nc\bfr{{\bf r}}\nc\bfR{{\bf R}}\nc\cR{{\mathcal R}}
\nc\bfs{{\bf s}}\nc\bfS{{\bf S}}\nc\cS{{\mathcal S}}
\nc\bft{{\bf t}}\nc\bfT{{\bf T}}\nc\cT{{\mathcal T}}
\nc\bfu{{\bf u}}\nc\bfU{{\bf U}}\nc\cU{{\mathcal U}}
\nc\bfv{{\bf v}}\nc\bfV{{\bf V}}\nc\cV{{\mathcal V}}
\nc\bfw{{\bf w}}\nc\bfW{{\bf W}}\nc\cW{{\mathcal W}}
\nc\bfx{{\bf x}}\nc\bfX{{\bf Z}}\nc\cX{{\mathcal X}}
\nc\bfy{{\bf y}}\nc\bfY{{\bf Y}}\nc\cY{{\mathcal Y}}
\nc\bfz{{\bf z}}\nc\bfZ{{\bf Z}}\nc\cZ{{\mathcal Z}}

\nc\sW{{\mathscr W}}
\nc\sC{{\mathscr C}}

\newcommand\complexes{{\mathbb C}}
\newcommand\ff{{\mathbb F}}

\newcommand\integers{{\mathbb Z}}
\nc\dgv{\delta_0}
\nc\rcrit{r_{c }}
\nc\bra[1]{\langle#1|}\nc\ket[1]{|#1\rangle}

\begin{document}

\title[Low-rate reliability bound]
{A low-rate bound on the reliability of a quantum discrete 
memoryless channel$^1$}
\author[Alexander Barg]{Alexander Barg}
\address{
Bell Labs, Lucent Technologies, 700 Mountain Avenue, 
Rm. 2C-375, Murray Hill, NJ 07974 USA}
\email{abarg@research.bell-labs.com}
\begin{abstract} We extend a low-rate improvement of the random coding bound
on the reliability of a classical discrete memoryless channel to its
quantum counterpart. The key observation that we make is that the problem
of bounding below the error exponent for a quantum channel 
relying on the class of stabilizer codes is equivalent to the problem
of deriving error exponents for a certain symmetric classical channel.
\end{abstract}
\maketitle
\footnotetext[1]{Research supported
in part by the Binational Science Foundation (USA-Israel), grant 
no. 1999099.}
\section{Introduction} 

Derivation of error bounds in quantum information theory 
is usually performed by translation of the standard methods
from its classical counterpart. Error exponents for the
classical-quantum channel (transmission of orthogonal states)
were derived in \cite{hol00}. Here we are concerned with the so-called
quantum-quantum channel which is the standard universe for quantum 
error-correcting codes.
An exponential upper bound on the distortion (error) probability was derived 
in a recent
paper \cite{ham01}. Here we show that this bound can be improved
for low noise and low values of the transmission rate. 
In Sect. \ref{sect:prelims} we give precise definitions of the quantum 
discrete memoryless channel (henceforth QDMC), codes, decoding, and 
error probability. Sect. \ref{sect:stab} contains
a brief review of stabilizer codes and their decoding. It turns out that
if we restrict
ourselves to the the class of stabilizer codes, then the bounds on their
distortion exponent also follow from the corresponding classical results.
In particular, in Sect. \ref{sect:rce} we give a short proof  
of the result of \cite{ham01}. The link to the classical results motivates 
us to derive a low-rate 
error exponent for a QDMC (Sect. \ref{sect:x}). A condition when it 
improves the random coding bound of \cite{ham01} is given. 
We conclude by specializing the results to the case of a depolarizing 
channel and showing a
concrete improvement for low code rates in the case of low noise.

\section{Preliminaries}\label{sect:prelims}

A quantum $d$-ary digit, a {\em qudit}, is a $d$-dimensional complex
space $H=\complexes^d,$ where $d$ will be assumed a prime number. 
Below by $\cX$ we denote the finite field $\ff_{q},$ where $q=d^2.$ 
We consider transmission of unit-length state vectors $\ket\psi$ from the  
$d^n$-dimensional space $H_n=H^{\otimes n}.$ Let us fix some orthonormal 
basis of $H$ and write it as $(\ket 0,\ket 1,\dots,\ket {d-1}).$
A unitary basis of error operators (an error basis, for short) is
defined as $\{E_{i,j}=X^iZ^j, i,j\in \ff_d\},$  
\[
X\ket i=\ket{(i-1)\text{mod } d},\quad Z\ket j=\omega^j\ket j,
\]
and $\omega$ is a primitive $d$th root of unity.

A {\em quantum discrete memoryless channel}
$\sW$ is defined as an arbitrary collection of operators of the form 
$(A_{u}, u\in \cX)$, where 
\[
A_u=\sum_{v\in \cX} a_{uv}E_v
\]
and where the complex row vectors $a_v=(a_{uv}, u\in \cX)$ define a 
probability distribution on $\cX$ given by
\[
W(v)=a_v a_v^\ast \quad (v\in \cX), \quad \sum_v W(v)=1.
\]
We note that this definition is derived from the general definition
of the quantum channel $\Phi$ which is a trace-preserving 
completely positive map on the set of density operators on $H_n$.
It is known that any such map 
$\Phi$ can be written as
\[
\Phi(S)=\sum_k A_k S A_k^\ast
\]
for some set of operators $A_k$, where $S$ is a density operator
on $H_n$ (the so-called Kraus representation of the channel). 
The absence of memory in the channel is
reflected by the fact that the operators $A_k$ can be written as
tensor products of operators on $H.$

As an example, let $d=2$ and consider the so-called 
{\em depolarizing channel} 
$\sW=\{\sqrt{1-p}I,\sqrt{p/3}\sigma_x,\linebreak[2]
\sqrt{p/3}\sigma_z,\sqrt{p/3}\sigma_y\},
$ where $(\sigma_x,\sigma_z,\sigma_y)$ is the set of Pauli matrices.
This channel acts on qubits by phase flips, amplitude flips, or
combinations of both applied with probability $p/3$ each.
More generally, for any $d$ we can define a depolarizing channel
as follows: $\sW=\{\sqrt{1-p}I;\sqrt\frac p{q-1}E_{i,j}, i,j\in \ff_d\}.$

A quantum code $\cQ$ is a linear subspace of $H_n.$ The {rate} of $\cQ$
is defined as $R=R(\cQ):=(\log_dK)/n,$ 
where $K$ is the dimension of $\cQ.$ Let $\cR$
be a recovery operator, i.e., another completely positive trace-preserving
map on $H_n$, restricted to $\cQ$. The {\em fidelity} of the code $\cQ$
for a given channel $\Phi$ and a given recovery operator $\cR$ equals
\[
F(\cQ,\{\Phi,\cR\})=\frac{1}{K}\min_{B\subset \cQ}
\sum_{\psi\in B} \bra\psi\cR\Phi [\ket\psi\bra\psi]\ket\psi,
\]
where the minimum is taken over all orthonormal bases $B$ of the code.
In particular, for the QDMC defined above, $\Phi=\sW^{\otimes n}.$ Below
we will omit the recovery operator from the notation.

For a given rate $R$ we wish to define the reliability (exponent) 
of a QDMC $\sW.$ Let 
\[
E(n, R,\sW)=\sup_{\cQ\subset H_n: R(\cQ)\ge R}-\frac 1n \log_d(1-F(\cQ,\sW))
\]
be the error exponent for the rate $R$ and code length $n$.  Let
\[
E(R,\sW)=\liminf_{n\to\infty} E(n,R,\sW).
\]

Let $H_m(Q)=-\sum_{x\in\cX}Q(x)\log_mQ(x)$ be the  entropy of a 
probability distribution $Q$ on $\cX$. 
For two probability distributions $P$ and $Q$, their information
divergence is given by 
$D_m(Q\|P)=\sum_{x\in \cX}Q(x)\log_m\frac{Q(x)}{P(x)}$ (if the base of
the logarithms and exponents below is omitted, it is equal to $d$).

The following theorem was proved in \cite{ham01}.
\begin{theorem}\label{thm:qrce} \cite{ham01}
For any rate $R\ge 0$ and any QDMC $\sW$
\begin{equation}\label{eq:qrce}
E(R,\sW)\ge E_r(R,\sW)=\min_{V}[D(V\|W)+|1-H(V)-R|^+],
\end{equation}
where the minimum is taken with respect to all probability distributions
on $\cX$ and $|a|^+:=\max(a,0).$ 
\end{theorem}
Since $E_r(R,\sW)>0$ for $0\le R< 1-H(W),$ this result also implies
a lower bound of $1-H(W)$ on the capacity of the channel $\sW$.

Given a vector $x\in \cX^n,$ we can define an empirical probability 
distribution $P$ on $\cX$ given by $P(u)=|\{i: x_i=u\}|/n, u\in \cX$.
Below we call it the {\em type} of the vector $x$ and write 
$T(x)=P.$ The type of the all-zero vector will be denoted by $P_0$;
we have $P_0(u)=\delta_{u,0}.$ The set of all sequences of a given
type $P$ will be denoted as ${\sf T}_P(\cX^n).$ It is clear that
\[
|{\sf T}_P(\cX^n)|= \exp_q(n (H_q(P)+o(1))).
\]
Let $\cP(\cX^n)$ be the set of all types on $\cX^n.$ Obviously,
\[
|\cP(\cX^n)|=\binom{n+q-1}{q-1}\le n^q \quad(n, q\ge 2).
\]
For any $x\in \cX^n$ and any stochastic matrix $V: \cX\to \cY$, 
the $V$-{\em shell} of $x$ is defined as the set ${\sf T}_V(x)\subset \cY^n$ 
formed by those $y$ whose conditional type is $V$. This means that for
any such $y$ its type is $T(y)=PV,$ where $PV$ is the probability distribution
on $\cY$ given by $PV(y)=\sum_{x\in\cX} P(x)V(y|x).$

\section{Stabilizer codes and their decoding}\label{sect:stab}
The construction of stabilizer quantum codes in \cite{cal97b}, \cite{cal98}
is as follows. Consider the vector space $V_n=(\ff_d\times\ff_d)^n.$ 
Write a typical vector $x\in V_n$ as $(x_1,x_1',x_2,x_2',\dots,x_n,x_n')$  and 
consider a standard symplectic form on $V_n$ defined by
\[
( x,y )=\sum_{i=1}^n x_iy_i'-x_i'y_i.
\]
Now let $\cC\subset \cX^n$ be an additive code, i.e., an additive subgroup 
of $\cX^n$ and define $\cC^\bot$ as the set of vectors in
$(\ff_q^+)^n\cong V_n$ that are $(\,,\,)$-orthogonal
to every vector in $\cC.$ Suppose that the number
of vectors in $\cC$ is $q^k$ so that the rate of $\cC$ equals
$R(\cC)=k/n.$ We then have $|\cC^\bot|=q^{n-k}.$

We begin with a pair of codes $\cC^\bot\subset \cC\subset \cX^n$
and a set $\cE\subset \cX^n$ such that
\[
\forall_{x,y\in \cE} (y-x\in \cC) \;\Rightarrow\; (x=y).
\]
According to this definition, we can take at most one error
vector per coset of $\cX^n/\cC$ and therefore, the maximum size
of the set $\cE$ equals $q^{n-k}.$
It is possible to construct a quantum code $\cQ\subset H_n$
of (complex) dimension $d^{2k-n}$ which is an invariant subspace
of the set of error operators $N_\cE=\{N_x, x\in \cE\}$ given by
\[
N_x=\bigotimes\limits_{i=1}^n N_{x_i},
\]
where for every $i$ the operator $N_{x_i}=E_{x_{i,1},x_{i,2}}$ is an 
element of the error basis determined by the representation of the 
coordinate $x_i\in \ff_q$ of $x$ as a pair of elements 
$(x_{i,1},x_{i,2})\in (F_d)^2.$
Moreover, there are $d^{2(n-k)}$ such invariant subspaces whose
orthogonal direct sum equals $H_n.$ Thus the rate $R$ of the stabilizer
code $\cQ$ is related to the rate of $\cC$ as $R=2R(\cC)-1.$

A stabilizer code $\cQ$ is $\cE$-error-correcting in the sense
that the action of any error operator from the set $N_\cE$ 
can be removed from the transmitted state. The received state $w$
is measured with respect to the set of pairwise orthogonal
operators $P_i$, each being an orthogonal projector on the
subspace of $H_n$ that corresponds to a coset of $\cX^n/\cC.$
Then within this coset we find one of the most probable error
vectors and recover the transmitted state by applying the inverse
error operator. 

The following bound on the fidelity of a given stabilizer code $\cQ$
was proved in \cite{ham01} based on a result in \cite{pre99}.
\begin{theorem} \label{thm:fidelity}\cite{ham01}
Let $\cQ$ be an $\cE$-error-correcting stabilizer
code used over a QDMC $\sW$. Then
\[
1-F(\cQ,\sW)\le \sum_{x\not\in \cE} W^n(x).
\]
\end{theorem}
This theorem provides a link between the quantum and the classical setting
which will be pivotal in our argument.

Note that there is substantial freedom in the choice of the error
set $\cE.$ To derive our result, we will take $\cE$ as follows. 
As pointed out above, the channel $\sW$ defines a probability
distribution $W$ on $\cX.$ For an additive code $\cC$ consider the quotient 
space $\cX^n/\cC.$
From each coset $S$ we take  one of the vectors $y=y(S)$ whose
probability $W^n(y)=\prod W(y_i)$ is the largest in $S.$ 
Finally, we take $\cE=\cup_{_S} y(S).$ 

We conclude this section by deriving a general analog of the weight
distribution and of the Gilbert-Varshamov bound for additive codes over $\cX$. 
For $q=4$ and the Hamming weight distribution
this result was proved in \cite{ash99i}.

\begin{theorem}\label{thm:gv} 
For any rate $R(\cC)>0$ and any $\delta>0,$  there 
exists an additive code $\cC\subset \cX^n$ of size $\exp(nR(\cC))$ such that 
$\cC^\bot\subset \cC$ and
for any type $P\ne P_0$,
\begin{equation}\label{eq:additive}
|\cC\cap {\sf T}_P|\le \exp_q[n(R(\cC)+H_q(P)-1+\delta)].
\end{equation}
In particular, for any $x\in \cC\backslash\{0\}$ with $T(x)=P$ we have
\[
R(\cC)\ge 1-H_q(P)-\delta.
\]
\end{theorem}
\Proof Let
\[
S_{n,k}=\{\cC\in \cX^n: \log_q|\cC|=k,
\cC^\bot\subset \cC\}.
\]
It was observed several times in the literature (e.g., \cite{cal97b}, 
\cite{ham01}) 
that every vector $x\in \cX^n\backslash\{0\}$ is contained in the same
number of codes in $S_{n,k}.$ Denote this number by $B$ and let $S=|S_{n,k}|.$ 
Counting in two ways the sum of sizes of all the codes in $S_{n,k}$
we obtain
\(
(q^n-1)B=(q^k-1)S.
\)
Let us fix a type $P$. Clearly,
\[
\sum_{P'\in \cP^n(\cX):\; H_q(P')\le H_q(P)} |{\sf T}_{P'}|\le n^q q^{nH_q(P)}.
\]
Thus as long as $ n^q q^{nH_q(P)}B< S$ or
\[
n^q q^{nH_q(P)}<\frac{q^n-1}{q^k-1}=q^{n(1-R(\cC))}(1+o(1)),
\]
there exists a code $\cC\in S_{n,k}$ such that for every $x\in \cC\backslash
\{0\}$ we have $H_q(T(x))\ge H_q(P).$ This proves the last part of the claim.

For any $P\ne P_0$ the average number of code vectors of type $P$ in a code
$\cC\in S_{n,k}$ equals 
\[
\frac1S\sum_{{\cC}\in S_{n,k}}|\{x\in (\cC\cap {\sf T}_P)\}|= 
\frac{B|{\sf T}_P| }{S}= \exp_q[n(R(\cC)+H_q(P)-1+o(1))].
\]
Since there are no more than $n^q$ different types, this proves
the first part of the claim. \qed

\section{The random coding bound}\label{sect:rce}

Let $\cX$ be an input and $\cY$ an output alphabet of a classical DMC
given by a stochastic matrix $W(y|x)$. Suppose that $\cX\subset \cY$ and that
$\cY$ is an abelian group, written additively. A channel
is called {\em additive} if $W(y|x)$ depends only on the difference $y-x,$
i.e., $W(y|x)=W(y-x)$ (the last term is actually $W(y-x|0),$ but below
we abuse the notation slightly and use unconditional distributions).
Note that an additive channel $W$ is symmetric in 
the sense that every row is a permutation of a fixed probability 
vector, and the same is true with respect to every column. 
By Theorem \ref{thm:fidelity} the problem of bounding from below
the reliability exponent of a QDMC is now {\em reduced to the corresponding 
classical problem for a symmetric, additive DMC} with $\cY=\cX.$ 
With this observation Theorem \ref{thm:qrce} follows by a combination
of standard arguments; so having in mind the reader well familiar
with error exponents of classical channels we could as well stop here.
In the interest of staying self-contained we will supply some more details.

\medskip
A. {\sc General form of the random coding exponent.}
For any type $P\in \cP(\cX^n)$ and any stochastic 
$|\cX|\times|\cY|$ matrix $V$, let 
\[
D(V\|W|P)=\sum_{x,y}P(x)V(y|x)\log \frac{V(y|x)}{W(y|x)}
\]
be the conditional divergence and
\[
I(P,V)=\sum_{x,y}P(x)V(y|x)\log \frac{V(y|x)}{\sum_x P(x)V(y|x)}
\]
be the mutual information between $x\in {\sf T}_P(\cX^n)$ and 
$y\in {\sf T}_V(x)$. 
The following theorem (reformulated slightly from
\cite{csi98}) gives one of the general forms of the error exponent of a
classical DMC.
\begin{theorem}\label{thm:genrce} For a given type $P\in \cP(\cX^n)$ let 
$A\subset {\sf T}_P(\cX^n), |A|=d^{(R'-\epsilon)n}$ 
be a code such that for every stochastic matrix $\tilde V:\cX\to \cX$ 
\begin{equation}\label{eq:dd0}
|\{(x_i,x_j) \in A\times A: \; x_j\in {\sf T}_{\tilde V}(x_i)\}|
\le \exp[{n(R'-I(P,\tilde V))}].
\end{equation}
Suppose that $A$ is used over a DMC $W:\cX\to\cY$ 
together with a maximum mutual
information decoder. Then the exponent $E(A,W)$ of the maximum error
probability $\max_{x\in A} p_e$ satisfies $E(A,W)\ge E_r(P,R',W),$ 
where
\begin{equation}\label{eq:rce}
E_r(P,R',W)=\min_{V}[D(V\|W|P)+|I(P,V)-R'|^+],
\end{equation}
and where $V$ runs over the set of all channels $\cX\to \cY.$
\end{theorem}

{\em Remarks.}
1. This theorem is a generalization of a classical fact of coding
theory, that ``binary linear codes of rate $R$ and weight distribution
$A_w\le 2^{n(R-1)}\binom nw, w=1,2,\dots, n$ achieve the random coding
exponent of the binary symmetric channel.''

2. The best bound on the reliability exponent of the channel $W$
is obtained by computing the maximum on $P$ in (\ref{eq:rce}). The
quantity
$E(R',W)=\max_P E_r(P,R',W)$ is usually called the random coding
exponent of $W$.

3. The maximum mutual information decoder, which is used to prove this 
result and which was employed in \cite{ham01}, 
is different from the decoder defined in Sect. \ref{sect:stab}. 

\medskip
B. {\sc Additive channels and codes.} 
Recall that in our problem $\cX$ is an additive group and that $\cY=\cX$. 
Further, since the channel $W$ is symmetric, the maximizing input distribution
$P$ in (\ref{eq:rce}) is known to be uniform \cite{dob63}: 
$P_u(x)=|\cX|^{-1}$ for any $x\in \cX.$

Let us substitute $P_u$ into the condition (\ref{eq:dd0}) on the 
``distance distribution'' of the code $A.$
Let $x$ be a vector such that $T(x)=P_u$ and let $\tilde V$ be a 
stochastic matrix such that
${\sf T}_{\tilde V}(x)\cap {\sf T}_{P_u}(\cX^n)\ne\emptyset.$ 
Then for any letter $x\in \cX$ the sum $\sum_{xf\in\cX} \tilde V(y|x)=1.$ 
We compute
\[
I(P_u,\tilde V)=\log|\cX|-H(\tilde V|P_u),
\]
So the upper bound in (\ref{eq:dd0}) takes the form 
\begin{equation}\label{eq:dd01}
|\{(x_i,x_j) \in A\times A: \; x_j\in {\sf T}_{\tilde V}(x_i)\}|
\le \exp[n(R'+H(\tilde V|P_u)-\log|\cX|)].
\end{equation}

Now consider the code $\cC$ from Theorem \ref{thm:gv}. Almost all of its
codewords are of type $P_u$ and nearby types (types close to it in
some suitable metric, say, the $\ell_1$-distance). 
We claim that the ``distance distribution'' of the
code $\cC$ satisfies (\ref{eq:dd01}). Since the code is additive, 
it suffices to consider matrices $\tilde V$ such that $\tilde V(y|x)$ 
depends only on the difference $y-x.$ Any such matrix defines a distribution 
$\tilde V(z)=\tilde V(z|0)$ on $\cX.$ Using this in (\ref{eq:dd01}), we 
observe that this condition reduces to the condition (\ref{eq:additive}) 
satisfied by the ``weight'' distribution of $\cC.$ 
Now recall from \cite{csi81b} that the 
function $E_r(P,R',W)$ is uniformly continuous on $P$ and that, 
on account of the channel and code
being additive, the error probability of decoding does not depend
on the transmitted codeword. Therefore for growing $n$ the 
error exponent of the code $\cC$ attains the bound $E(R',W)$. 
This proves Theorem \ref{thm:qrce}.

Transforming the exponent (\ref{eq:rce}) to the form (\ref{eq:qrce})
is a matter of calculation. Indeed, let us substitute $P_u$ in (\ref{eq:rce}). 
Clearly, $D(V\|W|P)=D(V\|W),$ where on the right-hand 
side $V$ and $W$ are probability distributions on $\cX$ given by 
$W(z)=W(y|x), V(z)=V(y|x)$ for any $y,x$ such that $z=x-y.$ Further,
\begin{align*}
I(P,V)-R'=-|\cX|^{-1} \sum_{z\in \cX} H(V)+\log |\cX|-1-R=1-R-H(V),
\end{align*}
where we have used the relation $R'=2R(\cC)=1+R.$ 

\bigskip
{\sc Further observations.} 

\medskip
1. By the same token, the capacity of the quantum channel $\sW$ is
bounded below by the capacity of the classical symmetric channel $W.$ 
Again the
mutual information is maximized for the uniform input distribution,
which implies the bound $\sC\ge 1-H(W)$ independently of the results
on error exponents. Note however that when this result is specialized 
to the depolarizing channel (see Example in the next section), it falls 
below the best currently known estimate of \cite{div98}.

\medskip
2. If we return from (\ref{eq:rce}) to Gallager's original
form of the random coding bound (by a method outlined in 
\cite[pp.\,192-193]{csi81b}), 
the exponent (\ref{eq:qrce}) can be written in a somewhat more
convenient form. Namely:
\begin{theorem}\label{thm:qrceG}
Let $E_0(\rho,W)=\rho-(1+\rho)
\log
\sum_{x\in \cX} W(x)^{\frac{1}{1+\rho}}.$ Then
\[
E_r(R,\sW)=1-R-\log\Big(\sum_{x\in \cX}\sqrt{W(x)}\Big)^2 \quad(0\le R<\tfrac{\partial E_0}{\partial \rho}|_{\rho=1})
\]
and
\[
E_r(R,\sW)=\max_{0\le\rho\le 1}[-\rho R+E_0(\rho,W)]\quad(\tfrac{\partial E_0}{\partial \rho}|_{\rho=1}\le R\le 1-H(W)).
\]
\end{theorem}

\medskip 3.
In the classical setting, the line of thought realized
in Theorem \ref{thm:qrce} would correspond to an
attempt to prove error bounds for a general DMC relying on the class of 
additive codes.
It is well known \cite{dob63},\,\cite{csi81b} that this approach produces good
results only when the optimizing probability distribution on the
input alphabet is uniform. The classical channel
derived from a general QDMC for stabilizer codes turns out to be
{additive} and hence symmetric. Hence the lower bounds
on the reliability exponent thus obtained are arguably rather strong.

\section{Expurgation exponent for a QDMC}\label{sect:x}

Let $\cQ\subset H_n$ be a stabilizer code of rate $R=R(\cQ)$ used
over a QDMC $\sW$ together with the decoder defined in Sect. \ref{sect:stab}.
Define the $W$-weight of a letter $x\in\cX$ as
\[
|x|_{_W}=-\log\sum_{e\in \cX}\sqrt{W(e)W(e-x)},
\]
where $\log 0=-\infty$ by definition.

\begin{theorem}\label{thm:ex}
\[
E(R,\sW)\ge E_x(R,\sW)=\min_{P:\, H(P)\ge 1-R}\;
\Big[ \sum_{x\in \cX}P(x) |x|_{_W} -(R+H(P)-1)\Big].
\]
\end{theorem}
\Proof We start with the code $\cC$ whose existence is proved in Theorem
\ref{thm:gv}. Let $\cQ$ be the stabilizer quantum code associated with it.
By Theorem \ref{thm:fidelity}
\begin{align*}
1-F(\cQ,\sW)&= \sum_{e\not\in\cE} W^n(e)=\sum_{x\in \cC\setminus\{0\}} 
\sum_{\genfrac{}{}{0pt}{}{y\in \cX^n}{ W^n(y-x)\ge W^n(y)}} W^n(y)\\[2mm]
&\le \sum_{x\in \cC\setminus\{0\}}\sum_{y\in\cX^n}\sqrt{W^n(y)W^n(y-x)}
\end{align*}
\begin{align*}
&=\sum_{P\in \cP(X^n)}\sum_{x\in \cC\cap{{\sf T}_p(\cX^n)}}
\sum_y\sqrt{W^n(y)W^n(y-x)}\\[2mm]
&\le\sum_{P\in \cP(X^n)} \exp_d[2n(R(\cC)+H_q(P)-1+o(1))-n\sum_{x\in\cX}
P(x)|x|_{_W}],
\end{align*}
where the last step follows because the channel is memoryless. Conclude
by computing the logarithm and substituting the relation $2R(\cC)=1+R.$
\qed

\bigskip
Note that it is possible that $E_x(R,\sW)$ becomes infinite for 
$R\downarrow R_\infty(\sW)>0,$ which means that for rates $R<R_\infty(\sW)$
errors outside the set $\cE$ occur with probability zero. The quantity 
$R_\infty(\sW)$ gives a lower bound on the 
zero-error capacity of the channel $\sW$. 
Shannon's classical example of a channel with $R_\infty(\sW)>0$
\cite[p.\,532]{gal68} 
is given by the additive channel with $\cX=\integers_5$ and 
$W(x)=W(x+1)=1/2.$ Clearly, $R_\infty(\sW)>0$ if and only if 
$|x|_{_W}=0$ for some $x\in \cX$. A channel is
called {\em indivisible} if this condition does not hold, and hence
$R_\infty(\sW)=0.$

The function $E_x$ can be transformed to a different form, also due
to Gallager \cite{gal68}:
\[
E_x(R,\sW)=\sup_{\rho\ge 1}[-\rho R+E_{ex}(\rho,W)],
\]
where
\[
E_{ex}(\rho,W)=-\rho\log_d\frac1{d^2} \sum_{x\in \cX}
\Big(\sum_{e\in \cX}\sqrt{W(e)W(e+x)}\Big)^{1/\rho}.
\]

Let us state a condition for the bound $E_{ex}(R,\sW)$ to improve
the result of Theorem \ref{thm:qrce}. As remarked above, the optimizing
probability distribution on $\cX$ for the random coding bound 
(\ref{eq:rce}) in our case is uniform. Moreover, the exponent $E_x$
is also derived under the same assumption. It is known \cite{gal68}
that for one and the same input distribution and for code rates
$R< \partial E_{ex}(\rho,W)/\partial\rho\vert_{\rho=1}$
the function $E_{x}(R,\sW)$ is greater than $E_r(R,\sW)$,
so in this region of rates Theorem \ref{thm:ex} improves the result of 
Theorem \ref{thm:qrce}. 
Hence if the point $R_x=\partial E_{ex}(\rho,W)/\partial\rho\vert_{\rho=1}>0$
then there is a nonempty interval of code rates where 
$E_x(R,\sW)>E_r(R,\sW).$ Note that typically
such an interval exists only for low noise level in the channel.
To make an analogy with the classical case, the improvement takes place
if the value of the code rate $R(\cC)$ that corresponds to $R_x$ is greater
than $1/2$. In the range where it improves the bound (\ref{eq:qrce}),
the exponent $E_x(R,\sW)$ can be written as
\begin{equation}\label{eq:exlow}
E_x(R,\sW)=\min_{P:H(P)=1-R} {\sf E}\, |X|_{_W},
\end{equation}
where $X$ is a random variable on $\cX$ distributed according to $P.$
This follows by the Gilbert-Varshamov bound of Theorem \ref{thm:gv}.

\remove{
\begin{align*}
&-\frac 1q\sum_z\log\sum_e\sqrt{W(e)W(e+z)} +\log\frac 1q
\Big(\sum_x\sqrt{W(x)}\Big)^2\\
&=-\frac1q\sum_z\log\frac{\sum_e\sqrt{W(e)W(e+z)}}{\frac 1q
(\sum_x\sqrt{W(x)})^2}\\
&\ge\log\frac{\sum_z\sum_e\sqrt{W(e)W(e+z)}}{(\sum_x\sqrt{W(x)})^2}=0.
\end{align*}
Hence }

\bigskip
{\em Remark.} The general form of the function $E_x(R,\sW)$
for a given additive, indivisible channel $\sW$ is as follows:
\[
E_x(R,\sW)=\max_P\sup_{\rho\ge 1}[-\rho R+E_{ex}(\rho,P,\sW)],
\]
where
\[
E_{ex}(\rho,P,\sW)=-\rho\log_d \sum_{x,x'\in \cX}P(x)P(x')
\Big(\sum_{e\in \cX}\sqrt{W(x-e)W(x'-e)}\Big)^{1/\rho}.
\]
Optimization on the input distribution $P$ in this expression is easy 
if the $q\times q$ matrix
\[
[(\sum_{e\in \cX}\sqrt{W(x-e)W(x'-e)})^{1/\rho}]
\]
is nonnegative
definite for every $\rho\ge 1$ \cite{jel68}, and turns into a 
difficult problem otherwise. For the channel to be 
nonnegative definite it is sufficient that 
for every pair of distinct vectors $(x,x')$ the sum on $e$ in
the expression for $E_x(\rho,P,\sW)$ takes one and the same value 
(the so-called equidistant channels \cite{jel68}).
For equidistant channels the maximum on $P$ is achieved for the uniform
distribution $P(x)=1/q, x\in \cX.$ For instance, the $d$-ary depolarizing 
channel is equidistant. 
However, there are many examples of not nonnegative definite additive,
indivisible channels.  For instance, let $d=3.$ Consider
the channel given by the following probability distribution:
\begin{equation*}
\begin{array}{r*{9}c}
u &00&01&02&10&11&12&20&21&22\\
W(u)&0&0.49&0&0.01&0.01&0&0.49&0&0 
\end{array},
\end{equation*}
where $u\in (\ff_9)^+\cong\integers_3\times \integers_3.$
It is easily verified that this channel is not nonnegative definite for 
$\rho\ge 1.37.$ \qed

\begin{figure}[tH]
\begin{center}
\setlength{\unitlength}{1mm}
\begin{picture}(88,70)
\medskip\put(25,110){\epsfysize=80mm
\epsffile[72 400 840 720]{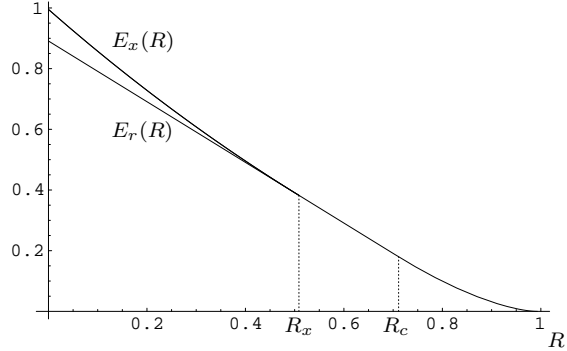}}
\put(45,24){\scriptsize $R_x$}
\put(58,24){\scriptsize $R_c$}
\put(80,22){\scriptsize $R$}
\put(22,62){\scriptsize $E_x(R)$}
\put(22,50){\scriptsize $E_r(R)$}
\end{picture}
\vskip-2cm
\begin{minipage}{\linewidth}
\caption{Error exponents for the depolarizing channel with 
$d=2$ and $p = 0.0005$.
For $0\le R\le R_x$ the function $E_x$ gives a stronger bound than 
$E_r$.}
\label{fig:QSC}
\end{minipage}
\end{center}
\end{figure}

\bigskip
{\em Example.} Let us specialize the results of Theorems \ref{thm:qrce} and
\ref{thm:ex} for the case of the $d$-ary depolarizing channel $\sW.$
Let us denote the reliability exponent of $\sW$ by $E(R,p).$
The result can be expressed in a closed form. Let 
\begin{eqnarray*}
h(x)&=&-x\log_q \frac{x}{q-1}-(1-x)\log_q x\\
D(x\|y)&=&x\log_q \frac{x}{y}+(1-x)\log_q \frac{1-x}{1-y}\\
\dgv(x)&=&h^{-1}(1-x).
\end{eqnarray*}
We have
\[
E(R(\cQ),\sW)\ge 2E_\ell((1+R)/2,p),
\]
where
\begin{align}
E_\ell(r,p)&=-\dgv(r)\log_q\gamma_q(p) &(0\le r\le r_x) \label{eq:depol_ex}\\
E_\ell(r,p)&=D(\rho_0\|p)+\rcrit-r &(r_x\le r\le \rcrit) \nonumber\\
E_\ell(r,p)&=D(\dgv(r)\|p) &(\rcrit\le r\le 1-h(p)), \nonumber
\end{align}
\[
r_x=1-h\Big(\rho_0\Big(2-\frac{q\rho_0}{q-1}\Big)\Big),
\quad \rcrit=1-h(\rho_0),
\]
\[
\rho_0=\frac{\sqrt{p(q-1)}}{\sqrt{p(q-1)}+\sqrt{1-p}},\qquad
\gamma_q(p)=p\frac{q-2}{q-1}+2\sqrt\frac{p(1-p)}{q-1}.
\]

This reliability exponent can be obtained from Theorems 
\ref{thm:qrceG}, \ref{thm:ex} or computed directly 
starting with codes whose existence is proved in Theorem \ref{thm:gv}.
The expurgation exponent (\ref{eq:depol_ex}) is straightforward from 
(\ref{eq:exlow}). If $R_x:=2r_x-1>0,$ then from (\ref{eq:depol_ex})
we obtain an improvement over the result of 
Theorem \ref{thm:qrce} in the interval of values of $R$ between zero 
and $R_x.$ It turns out that this condition is
satisfied for low noise level (see an example in Fig. \ref{fig:QSC}).
For $d=2$ the expurgation bound improves the random coding
exponent for $0<p\le 0.004.$\qed

\medskip
{\em Acknowledgment.} The author is grateful to A. Ashikhmin and 
G. Kramer for helpful discussions.

\renewcommand\baselinestretch{0.9}
{\footnotesize
\providecommand{\bysame}{\leavevmode\hbox to3em{\hrulefill}\thinspace}

}

\end{document}